\documentclass[superscriptaddress,prd,aps,preprint,eqsecnum,floats]{revtex4}
\usepackage{epsfig}
\begin{document}
\title{Light meson orbital excitations in the QCD string approach}
\author{A.M. Badalian}
\affiliation{State Research Center,\\ Institute of Theoretical and Experimental
Physics, Moscow, Russia}
\author{B.L.G. Bakker}
\affiliation{Department of Physics and Astronomy, Vrije Universiteit, Amsterdam}
\begin{abstract}
In the framework of the QCD string approach it is shown that the
spin-averaged masses  $\bar{M}(nL)$ of all low-lying light mesons are
well described using the string tension $\sigma$ as the only
parameter.  The Regge slope $\alpha'_L$ and the intercept $\alpha_L(0)$
of the Regge $L$-trajectory for $\bar{M}(nL)$ are calculated
analytically and turn out to be $\alpha'_L = 0.80$ GeV${}^{-2}$ (for $L
\leq 4$) and $\alpha_L(0) = -0.34$, in good agreement with the
experimental data: $\alpha'_{L\,{\rm exp}} = 0.81 \pm 0.01$
GeV${}^{-2}$, $\alpha_{L\,{\rm exp}}(0) = -0.30 \pm 0.02$.  To obtain
this strong agreement with the data the nonperturbative quark
self-energy contributions to the meson masses must be taken into
account, which appeared to be large and negative for small values of
$L$, and are even for larger values of $L$ important for a close fit.
From the present analysis of the meson spectra the restriction
$\alpha_s \leq 0.40$ on the strong coupling is required.


\end{abstract}
\maketitle
\section{Introduction}
\label{sect.1.0}

The spectra of hadrons form an extremely important test ground for
nonperturbative QCD. The scaling property of QCD tells us that in the
end all characteristics of hadrons must depend on a single parameter,
say $\sigma$ or $\Lambda_{\rm QCD}$. Till now, all calculations of
hadronic spectra with an accuracy comparable to the uncertainties in
the experimental data, have relied on models that contain several, in
some cases even many, parameters.  Here we adopt the formalism that
uses the QCD-string Hamiltonian which relies on only one parameter--the
string tension $\sigma$, while for light mesons the interactions
derived from perturbative QCD, both static and spin-dependent, can be
considered as a perturbation.  The string tension can be extracted from
experiment, in particular from the slope of the leading Regge
trajectory, and in the present paper we use it to describe the orbital
excitations of light mesons.

The QCD string approach  developed in recent years \cite{ref.1} -
\cite{ref.3}  starts from first principles, i.e. from the Euclidian QCD
Lagrangian. In Ref.~\cite{ref.2} the relativistic Hamiltonian $H_R$ for
the light  mesons with spinless quarks was derived under several
verifiable assumptions.

First, string excitations (hybrids) are not taken into account, since
there is a large gap, $\sim 1$ GeV, between a mesonic ground state and
its gluonic excitation \cite{ref.4}. Without this approximation one
obtains a matrix multichannel Hamiltonian \cite{ref.5}. Therefore the
ground states of the mesons with not too large orbital momentum ($ L\leq 5$)
can be treated in the closed channel approximation.

Secondly, the relativistic Hamiltonian $H_R$ used here is derived in
the quenched approximation, where creation of $q\bar{q}$ pairs (sea
quarks) is neglected. The accuracy of this approximation for low-lying
states is expected to be approximately 10\% \cite{ref.1}, while high
radial and orbital excitations can be strongly affected by $q\bar{q}$
pair creation\cite{ref.6}.

Thirdly, to derive the Hamiltonian $H_{\rm R}$ only forward-in-time quark
trajectories have been taken into account while backward trajectories
of a quark (antiquark) were neglected. The accuracy of this
approximation has been checked comparing meson spectra of the
center-of-mass Hamiltonian $H_{\rm R}$ and light-cone Hamiltonian
$H_{\rm LC}$, since in the latter case backward-in-time trajectories do
not contribute \cite{ref.7}.  This comparison has shown that the
differences in meson masses for these Hamiltonians is not larger than
10\% for all mesons with exception of the pion, which receives a large
contribution from backward-in-time trajectories corresponding to
negative energy components and for them the formalism should be
modified. In the Hamiltonian used chiral effects are not taken into
account.

Under these assumptions and with the use of the
Fock-Feynman-Schwinger (FFS) representation the Green's function $G$ of a
meson with a spinless quark and antiquark can be written in a gauge
invariant way as a functional integral with the action $A$ \cite{ref.2}:
\begin{eqnarray}
 G & = & \int^\infty_0 d s\,\int^\infty_0 d \bar{s}
 \;Dz D\bar{z} \;\exp(-A), \nonumber \\
 A & = & K  +  \bar{K} + \sigma S_{\rm min},
\label{eq.1a}
\end{eqnarray}
where in the action (\ref{eq.1a}) the only approximation made is that
the vacuum average over the Wilson loop $\langle W(C) \rangle$ is taken
in the form of the minimal area law, viz
\begin{equation}
  \langle W(C) \rangle = const \; \exp ( - \sigma S_{\rm min}) .
\label{eq.2a}
\end{equation}
The accuracy of this approximation is determined by the condition
\begin{equation}
 R \gg T_g , 
\label{eq.3a}
\end{equation}
where $R$ is the size (e.g. the r.m.s. radius) of the meson and $T_g$
is the gluonic (vacuum) correlation length  which determines how the
vacuum correlators decrease as a function of the separation $r$ between
the quark and the antiquark. The value of $T_g$ was calculated on the
lattice \cite{ref.8} and in the quenched case was found equal to $\sim
0.15-0.20$ fm, i.e. it is much smaller than the r.m.s. radii $R(nL)$
for light mesons having $R(nL) \geq 0.8$ fm. Note that the condition
(\ref{eq.3a}) is also valid for most  excited heavy-light mesons and
even in heavy quarkonia, where e.g. $R(1P, c\bar{c}) \sim R(2P,
b\bar{b}) \sim 0.65$ fm.

In the general case the approximate area law can be replaced by the
exact $q\bar{q}$ interaction which contains a linear confining part for
distances satisfying Eq.~(\ref{eq.3a}) but should be modified at
smaller distances, being fully determined by a bilocal field
correlator. In the action $A$ (\ref{eq.1a}) $K$ and $\bar{K}$ are the
quark (antiquark) kinetic energy term written in the FFS
representation:
\begin{equation}
 K  = 
 \int_0^s \left[m^2 + \frac{1}{4}\dot{z}(\tau)^2_{\alpha} \right]\,d \tau,
 \quad \bar{K}  = 
 \int_0^s \left[m^2 + \frac{1}{4}\dot{\bar{z}}(\tau)^2_{\alpha} \right]\,d \tau,
\label{eq.4a}
\end{equation}
where $m$ is the current mass of a quark, $\tau$ is the proper time
introduced by Schwinger\cite{ref.9}, $z(\tau)$ and $\bar{z}(\tau)$ are
the paths of the quark and antiquark.

To define the Hamiltonian one can use the connection between the meson
Green's function and the Hamiltonian $H$:
\begin{equation}
 \partial G/ \partial T = - H_{\rm R} G ,
\label{eq.5a}
\end{equation}
where the Hamiltonian can be determined on any hypersurface, i.e. $H$
can be derived in different frames. Here we shall use the Hamiltonian
obtained in the c.m. frame, while in Ref.~\cite{ref.7} the Hamiltonian
was derived in the light-cone frame.

In order to use the relation (\ref{eq.5a}) in Euclidean space-time it
is of great importance to go over from the proper time $\tau$ to the
actual time $t \equiv z_4$, of a quark (antiquark). Doing so the new
quantity $\mu(t)$ is introduced:
\begin{equation}
 2 \mu(t) = \partial t / \partial \tau.
\label{eq.6a}
\end{equation}

After perfoming the canonical quantization the variable $\mu$ (being a
canonical coordinate) will define the constituent mass of a quark. The
last term in the action (\ref{eq.1a}) has the form of the Nambu-Goto string:
\begin{equation}
 S_{\rm min} = \int_0^T dt \,\int_0^1d\beta \,\sqrt{\det g},
\label{eq.7a}
\end{equation}
where  $g_{ab} = \partial_a w_\alpha \partial_b w^\alpha$  with 
$a,b = t,\beta$ and $w_\alpha (t,\beta)$ are the coordinates of the
string world surface. In Refs.~\cite{ref.2} the string was approximated
by a straight line connecting the path coordinates $z_\alpha (t)$
and $\bar{z}_\alpha (t)$ and in this case
\begin{equation}
 w_\alpha (t,\beta)=z_\alpha (t)\beta + \bar{z}_\alpha (t)(1-\beta),
 \quad   0 \leq \beta \leq 1.
\label{eq.8a}
\end{equation}
Due to the presence of the square root in $S_{\rm min}$ in
Eq.~(\ref{eq.7a}) this term cannot be quantized and to get rid of it one
can introduce two auxiliary fields, $\nu(t,\beta)$ and $\eta(t,\beta)$
in the standard way as done in string theory\cite{ref.10}.

By definition the introduction of the auxiliary fields is accompanied
by the additional integration over $D\mu$, $D\nu$, $D\eta$ in the
functional integral defining the meson Green's function,
\begin{equation}
  G = \int \, D\mu D\nu D\eta DR_\alpha Dr_\alpha \; \exp(-A),
\label{eq.9a}
\end{equation}
where ``the center-of-mass'' coordinate $R_\alpha$ and ``the relative''
coordinate $r_\alpha$ are introduced instead of the path coordinates
$z_\alpha$ and $\bar{z}_\alpha$.

As shown in Refs.~\cite{ref.2} the integrations over $D\eta$ and
$DR_\alpha$ can analytically be performed in the integral
(\ref{eq.9a}) and after that the Green's function has a simpler form,
viz
\begin{equation}
 G = \int \,D\mu D\nu Dr \; \exp(-A_R).
\label{eq.10a}
\end{equation}

As the next step instead of performing the integration over $D\mu$
and $D\nu$ in the integral (\ref{eq.10a}) one can use an equivalent
procedure--to go over to the canonical quantization of the Hamiltonian
$H_{\rm R}$ which corresponds to the action $A_{\rm R}$. This
Hamiltonian in Minkowski space-time is obtained from $A_{\rm R}$ in a
standard way\cite{ref.2,ref.10} and is taken as a starting point in our
analysis (see Eq.~(\ref{eq.5})).

The quantization of $H_{\rm R}$ has been performed in two cases: the
quasiclassical quantization of $H_{\rm R}$ for large orbital momenta
$L, \; L \gg 1$, in Ref.~\cite{ref.11} and for not too large $L$ in
Refs.~\cite{ref.2}.  Here we are interested only in orbital excitations
with $L \leq 4$ and in this case the Hamiltonian $H_{\rm R}$ can be
presented as the sum
\begin{equation}
 H_{\rm R} = H_{\rm R}^{(1)} + \Delta H_{\rm str},
\label{eq.11a}
\end{equation}
where the term $\Delta H_{\rm str}$ is relatively small for $L \leq 4$
and can be considered as a perturbation. Then the problem reduces to
the quantization of more simple Hamiltonian $H_{\rm R}^{(1)}$ which
can be easily done (see Sections \ref{sect.3.0} and \ref{sect.4.0}). At
the final stage we obtain the surprising result that after quantization
the unperturbed Hamiltonian $\tilde{H}_{\rm R}^{(1)}$ coincides with the
Hamiltonian used in the relativistic potential model (RPM)
\cite{ref.12,ref.13}. In this way (for the states with $L \leq 4$) the
connection between the QCD string Hamiltonian $H_{\rm R}^{(1)}$ and the
RPM is established  and we can also calculate the corrections which are
absent in the RPM.

The first correction $\Delta_{\rm str}$, called the string correction,
comes from the term $\Delta H_{\rm str}$ in Eq.~(\ref{eq.11a}); it is
negative and varying from a value of about -50 MeV for $L=1$ to about 
-150MeV) for $L=4$.

A second correction to the meson mass, $\Delta_{\rm SE}$, is due to the
spin (colour magnetic moment) interaction of a quark (antiquark) with
the external (vacuum) field when the operator $\sigma_{\alpha\beta}
F_{\alpha\beta}$ is inserted in the Wilson loop. The form of this
operator is
\begin{equation}
 \sigma_{\alpha\beta} =
 \frac{1}{4}\;(\gamma_\alpha \gamma_\beta - \gamma_\beta \gamma_\alpha).
\label{eq.11b}
\end{equation}
This nonpertubative self-energy correction was analytically calculated
in Ref.~\cite{ref.14} with the the use of the FFS representation for the
quark Green's function. $\Delta_{\rm SE}$ is negative and has a rather
large magnitude (of the order of -400 MeV to -300 MeV) for all states
with $L \leq 4$, slightly decreasing with growing $L$. Owing to this
correction the correct value of the Regge intercept was obtained.

We use here the current quark mass $m=0$ and due to the procedure of
canonical quantization the constituent mass of a quark can be defined
in a rigorous way as the canonical coordinate $\mu$ which is equal to
the quark kinetic energy. It is of interest to note that the final
expression of the Hamiltonian $\tilde{H}_{\rm R}^{(1)}$ does not
contain the constituent mass at all.  However, the constituent mass
$\mu$ must be defined since it explicitly enters those terms in the
Hamiltonian, like the spin-dependent and self-energy terms, which are
considered as a perturbation.

Here we consider in detail the spin-averaged meson masses, $\bar{M}(nL)$,
or the centers of gravity of the $nL$-multiplets (i.e. neglect the
hyperfine and fine-structure splittings) for which the physical picture
is simpler and at the same time more universal since the
parameters do not depend on spin and isospin.

We concentrate mostly on the orbital excitations with $n = 0$ for which
experimental data exist for all ground states with  $L \leq 5$. Then
for the linear confining potential $\sigma r$ all meson masses
$\bar{M}(nL)$ can be expressed through a single parameter--the string
tension $\sigma$. The values of the slope and the intercept of the
Regge L-trajectory $(L \leq 4)$ will be calculated analytically: their
numerical values are $\alpha'_L = 0.80$ GeV${}^{-2}$ ($\sigma =0.18$
GeV${}^2$) and $\alpha_L(0) = -0.34$ turn out to be in very
good agreement with the experimental numbers. From the Regge slope a
restriction on the admissable values of the string tension follows:
$\sigma = 0.18 \pm 0.005$ GeV${}^2$ for the pure linear potential and
$\sigma = 0.19 \pm 0.01$ GeV${}^2$ if the Coulomb interaction is taken
into account.

The Coulomb contribution is mainly  important for the $1S$ and the $1P$ states
having values in the range $- 200$ to $- 100$ MeV, and is considered
here in a twofold way: from exact calculations with the linear plus
Coulomb potential and also when the Coulomb interaction is considered
as a perturbation; both considerations give very close results. For
$\sigma  = 0.19$ GeV${}^2$ the QCD coupling  is $\alpha_s = 0.39$ which
is typical for heavy quarkonia, and from our analysis of the meson
spectra the following restriction on the strong coupling $\alpha_s \leq
0.42$, is obtained. This number is in a good agreement with the
two-loop value of the freezing coupling constant obtained  in
background field theory \cite{ref.15}.

\section{Experimental data}
\label{sect.2.0}
	     
The experimental numbers for the spin-averaged masses $\bar{M}(L)$, or the
centers of gravity of the $1\, L$-multiplet (the ground states with $n=0$),
are presented in Table \ref{tab.1.0} and need some remarks.

First, all members of the $1\,{}^3P_J$ multiplet are supposed to be
known:  $a_2(1318)$, $a_1(1235)$, and the $a_0(980)$ too, are
considered to form the $1\,{}^3P_0$ multiplet with  $\bar{M}(1P) =
1252$ MeV.  Similarly, for the $f_J (1P)$ mesons the spin-averaged
mass is 1245 MeV \cite{ref.16}.

Second, in the case of $L=2$, the fine-structure splittings of the $1\,
D$-wave mesons are supposed to be suppressed as compared to the
$P$-wave states \cite{ref.17}. As a result, for all members of the
$1\, D_J$-multiplet, e.g. the $\rho_3(1.69)$ and $\pi_2(1.67)$, their
masses are very close to each other and one can expect that the true
value of $\bar{M}(1D)$ lies  between these two values. The same would
be  valid for the isoscalar mesons, if they were not mixed with other
hadronic states, and just this situation is observed in experiment
where the  masses of the $\omega_3 (1.67)$ and the $\omega(1.65)$ have
values close to the corresponding isovector mesons\cite{ref.16}.

Due to the suppression of the matrix elements (m.e.) like $<1/r^3 >$
the spin splittings for the higher orbital excitations like $1F$, $1G$,
etc. should be even smaller than for the $1D$ mesons. Therefore the
masses of the $a_4$(2.01) and the $f_4$(2.03) are supposed to be close
to $\bar{M}(1F)$ as well as the masses of the $\rho_5$(2.30) with $L =
4$, and $a_6$(2.45) and $f_6$(2.47) with $L = 5$, lie close to  their
centers of gravity.  The masses of all orbital excitations ($n=0$) can
be nicely described by the Regge $L$-trajectory (see Fig.1):

\begin{figure}
\begin{center}
\epsfig{figure=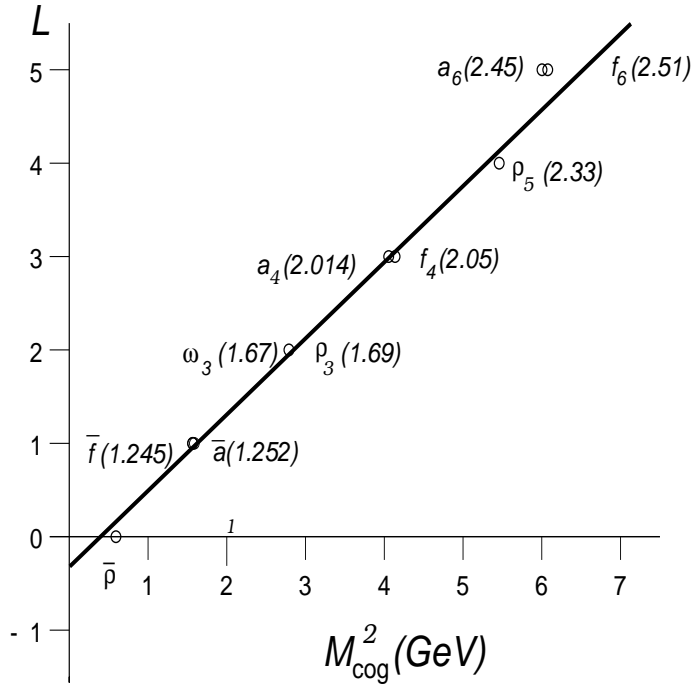,height=90mm,width=90mm}
\caption{The Regge $L$-trajectory for the light mesons. \label{fig.1}}
\end{center}
\end{figure}

\begin{equation}
  \bar{M}^2(L) = (1.23 \pm 0.02) \,L + 0.37 \pm 0.02\;  ({\rm GeV}^2),
 \label{eq.1} 
\end{equation} 
or 
\begin{equation}
  L  =  0.81 \bar{M}^2(L)  - 0.30
 \label{eq.2}
\end{equation} 
with the following Regge slope and intercept:  
\begin{equation}
 \alpha'_{L\,{\rm exp}} = 0.81 \pm 0.01 \; ({\rm GeV}^2) \quad {\rm  and} \;
 \alpha_{L\,{\rm exp}} (0) = -0.30 \pm 0.02\;  (L \leq  4).  
\label{eq.3}
\end{equation} 
and both values have a small experimental error.

Note that for the leading $\rho$-trajectory:

\begin{equation}
 J = \alpha'_J   M^2(J) + 0.48
 \label{eq.4} 
\end{equation}
the slope $\alpha'_J = 0.88$ GeV${}^{-2}$ and the intercept 
$\alpha_J(0) = 0.48$ are
larger since their values depend on the  spin contributions. On the
contrary, the $L$-trajectory is a universal one and in the approximation
of closed channels it is the same for isovector and isosinglet mesons.

In our paper the meson masses, Regge slope and Regge intercept will be
calculated analytically in the framework of the QCD string approach.

\section{Relativistic Hamiltonian} 
\label{sect.3.0}

We start with the Green's function Eq.~(\ref{eq.10a}) which is obtained
after performing two integrations $D \eta DR$ in the functional
integral, so in the Green's function still three integrations left:
over the two auxiliary fields $\mu(t)$ and $\nu(t)$ and also over the
separation $r$ between a quark and an antiquark, $r_\alpha = z_\alpha
(t) - \bar{z}_\alpha (t)$.  The explicit expression of this action
$A_{\rm R}$ was obtained in Refs.~\cite{ref.2} and the dependence of
$A_{\rm R}$ on $\mu$ and $\nu$ turns out to be rather complicated.
Therefore it is more simple and convenient to go over to the equivalent
procedure--canonical quantization of the Hamiltonian $H_{\rm R}$ which
corresponds to the action $A_{\rm R}$  in Eq.~(\ref{eq.10a}):
\begin{eqnarray}
  H_{\rm R} & = & \frac{p^2_r +m^2}{\mu(\tau)} + \mu(\tau) \nonumber \\
 & + & \frac{\vec{L}^{\, 2}}{r^2} \left[
 \mu(\tau) + 2 \int^1_0 d\beta \nu(\beta) (\beta - \frac{1}{2})^2
                            \right]^{-1} \nonumber \\
 & + &  \frac{1}{2} \sigma^2 r^2 \int^1_0 \frac{d\beta}{\nu(\beta)}
 + \frac{1}{2} \int^1_0 d\beta \nu(\beta]).
 \label{eq.5} 
\end{eqnarray}
By definition  the field operator $\mu(t)$ is
\begin{equation}
 \mu(t) =  \frac{1}{2} \frac{dt}{d\tau}, 
 \label{eq.6}
\end{equation} 
where $t$ is the actual time.

In Eq.~(\ref{eq.5}) $m$ is the current quark mass, which for a light
quark (antiquark) will be taken equal to zero; $\vec{L}$ is the angular
orbital momentum, $\vec{L} = \vec{r} \times \vec{p}$, and the operator
$p^2_r = (\vec{p} \cdot\vec{r})^2/(r^2)$. Canonical quantization of
$H_{\rm R}$ (\ref{eq.5}) has not been done in general, but was
performed in two cases: first when the orbital momentum $L \gg 1$ 
(the quasiclassical quantization in Ref.~\cite{ref.11}) and for relatively
small $L \leq 4$ which will be considered here. The constant $\sigma$
determining the nonperturbative potential is the string tension.

It is convenient to rewrite $H_{\rm R}$ as a sum of two terms:
\begin{equation}
 H_{\rm R} = H^{(1)}_{\rm R} + \Delta H_{\rm str},
 \label{eq.7} 
\end{equation} 
with the ``unperturbed'' Hamiltonian $H^{(1)}_{\rm R}$ defined by
\begin{equation}
 H^{(1)}_{\rm R} = \frac{\vec{p}^{\, 2} +m^2}{\mu(t)} + \mu(t) +
  \frac{1}{2} \sigma^2 r^2 \int^1_0 \frac{d\beta}{\nu(\beta)} +
 \frac{1}{2} \int^1_0 d\beta \nu(\beta).  
 \label{eq.8}
\end{equation} 
where in $H^{(1)}_{\rm R}$ we have included the term $L^2/(\mu r^2)$ and subtracted
the same term to give the string correction $\Delta H_{\rm str}$
\begin{eqnarray}
 \Delta H_{\rm str} & = & - \frac{\vec{L}^{\, 2}}{r^2} \left[
 \frac{1}{\mu(t)} - \frac{1}{\mu + 2 \int^1_0 d\beta \nu(\beta)
 (\beta - \frac{1}{2})^2} \right] \nonumber \\
 & = & -
 \frac{\vec{L}^{\, 2}}{\mu r^2}
  \frac{2 \int^1_0 d\beta \nu(\beta) (\beta - \frac{1}{2})^2}
       {\mu + 2 \int^1_0 d\beta \nu(\beta) (\beta - \frac{1}{2})^2}.
 \label{eq.9} 
\end{eqnarray}

If $L$ is not large, then the term $\Delta H_{\rm str}$ appears to be
relatively small and can be considered as a correction to the
Hamiltonian $H^{(1)}_{\rm R}$ \cite{ref.11} but for large $L$ the
representation of $H_{\rm R}$ as the sum Eq.~(\ref{eq.7}) is of no use,
since in this case both terms are equally important.  Note that to get
the expression (\ref{eq.8}) one needs the following definition
\begin{equation}
 \vec{p}^{\, 2} = p^2_r + \frac{\vec{L}^{\, 2}}{r^2}. 
 \label{eq.10}
\end{equation}
The simplest Hamiltonian  $H_0$ with $L = 0$ is  a special case of
$H_{\rm R}$ (or $H^{(1)}_{\rm R}$) with $\vec{p}^{\, 2}$ replaced by
$p^2_r$.

Since $\Delta H_{\rm str}$ is considered as a perturbation, the
canonical quantization needs to be performed only with the Hamiltonian
$H_{\rm R}^{(1)}$, which can be easily done, since the latter can be
presented as the sum of the kinetic energy operator $\hat{H}_{\rm kin}$
which depends on $\mu$ and $\hat{V}_{\rm pot}$ which depends on $\nu$:
\begin{equation}
 H_{\rm R}^{(1)} = \hat{H}_{\rm kin}(\mu) + \hat{V}_{\rm pot}(\nu).
\label{eq.3.7}
\end{equation}
The variables $\mu$ and $\nu$ are the canonical coordinates while the
momenta $\pi_\mu$ and $\pi_\nu$ canonically conjugated to $\mu$ and
$\nu$ turn out to be equal to zero, since $H_{\rm R}^{(1)}$ (as well as
$H_{\rm R}$) does not depend on the derivatives $\dot{\mu}$ and
$\dot{\nu}$. Hence it follows that
\begin{equation}
 \dot{\pi}_\mu = \partial H_{\rm R}^{(1)}/\partial \mu =0, \quad
 \dot{\pi}_\nu = \partial H_{\rm R}^{(1)}/\partial \nu =0.
\label{eq.3.8}
\end{equation}

Thus instead of calculating the Green's function as the functional
integral Eq.~(\ref{eq.10a}) and subsequently deriving the Hamiltonian
$\tilde{H}_{\rm R}^{(1)}$, one can obtain this Hamiltonian from $H_{\rm
R}^{(1)}$ Eq.~ (\ref{eq.8}) with the use of the extremum conditions
Eq.~(\ref{eq.3.8}). The equivalence of these two procedures was
explicitly illustrated in the third paper of Ref.~\cite{ref.2}
for the kinetic part in Eq.~(\ref{eq.3.7}):
\begin{equation}
 \hat{H}_{\rm kin}(\mu) = \frac{\vec{p}^{\,2} + m^2}{\mu^2}  + \mu(t).
\label{eq.3.9}
\end{equation}
For this Hamiltonian the explicitly calculated Green's function for the
free particle is
\begin{equation}
 G_0 = \int D\vec{z}(t) D\vec{p}\,
 \exp[i\int_0^T(\vec{p} \cdot \dot{\vec{z}}- \sqrt{\vec{p}^{\,2}+m^2})]
\label{eq.3.10}
\end{equation}
which is just the canonical representation of the Green's function with the
free Hamiltonian
\begin{equation}
 H_0 = \sqrt{\vec{p}^{\,2} + m^2}
\label{eq.3.10a}
\end{equation}

\section{The extremal values of the operators $\mu$ and $\nu$}
\label{sect.4.0}

To understand the physical meaning of the auxiliary fields $\mu(t)$ and
$\nu(t)$ let us find their extremal values. First, in the Hamiltonian
$H_{\rm R}$ we determine the variable $\nu(\beta)$ from the extremum
conditions (\ref{eq.3.8}:
\begin{equation}
 \frac{\delta H^{(1)}_{\rm R}}{\delta \nu(\beta)} = 0, \quad \frac{\delta
 H^{(1)}_{\rm R}}{\delta \mu(t)} = 0. 
 \label{eq.11} 
\end{equation}
Then one finds that $\nu(\beta)$, which is an operator in general,
does not depend on the string parameter $\beta$ and is equal to
\begin{equation}
 \nu_0(\beta)=\sigma r,
 \label{eq.12} 
\end{equation}
i.e. is actually  the energy density along the string.

With the use of Eq.~(\ref{eq.12}) the Hamiltonian $H^{(1)}_{\rm R}$ reduces to a
simpler operator:  
\begin{equation}
 H^{(1)}_{\rm R} =  \frac{\vec{p}^{\, 2} +m^2}{\mu(t)} + \mu(t) + \sigma r,
 \label{eq.13} 
\end{equation}
where $\mu(t)$ is still an operator in the Hamiltonian formalism. Its
extremum can be found from the second extremum condition (\ref{eq.11})
\begin{equation}
 \mu (t) = \sqrt{\vec{p}^{\, 2} + m^2}, {\rm for}\, H^{(1)}_{\rm R}, \quad
 \mu_0(t) = \sqrt{p^2_r + m^2}, {\rm for}\, H_0,
 \label{eq.14}
\end{equation}
i.e. the extremal value of $\mu$ is one half the kinetic energy
operator.  Note that after canonical quantization the `coordinate'
$\mu$ is already independent of time.  Substituting it into the
Hamiltonian $H^{(1)}_{\rm R}$ one obtains
\begin{equation}
 \tilde{H}^{(1)}_{\rm R} = 2\sqrt{\vec{p}^{\, 2} + m^2} + \sigma r
 \label{eq.15}
\end{equation} 
giving rise to an eigenvalue equation that is
identical to the spinless Salpeter equation (SSE) with a linear potential 
\begin{equation}
 \tilde{H}^{(1)}_{\rm R} \psi(nL) = M_0(nl) \psi(nL).  
\label{eq.16} 
\end{equation}
This equation has been used in the RPM for many
years\cite{ref.12,ref.13} and the only difference is that in the
Hamiltonian $\tilde{H}_{\rm R}^{(1)}$ in Eq.~(\ref{eq.16}) we must use
the current quark mass $m$ since just the current mass enters the meson
Green's function ($m=0$ for the light mesons) while in the RPM  $m \neq
0$ is usually used, e.g. in Ref.~\cite{ref.13} $m=(m_u + m_d)/2 = 220$
MeV.

\section{The constituent quark mass} 
\label{sect.5.0}

Although the constituent mass $\mu$ is not explicitly present in
$\tilde{H}^{(1)}_{\rm R}$, it enters many important physical characteristics like the
spin splittings and magnetic moments, and also in the string and
self-energy corrections, therefore it must not be left in as an
operator. The simplest way to solve this is to define $\mu_0$ as the
expectation value of one half the quark kinetic energy operator
Eq.~(\ref{eq.14}), i.e.,

\begin{equation}
  \mu_0 (nL) =  \langle \sqrt{\vec{p}{\, 2} + m^2} \rangle_{nL} .
 \label{eq.17} 
\end{equation}
Note that the eigenvalues $M_0(nL)$ in Eq.~(\ref{eq.15}) for the linear
potential $\sigma r$ are connected  with $\mu_0$ as follows
\begin{equation}
 M_0 (nL) = 4 \mu_0(nL).  
 \label{eq.18} 
 \end{equation}
The values of $\mu_0$ can be expressed through a single parameter--the
string tension $\sigma$ and the universal numbers $a(nL)$ given by
\begin{equation}
 \mu_0 (nL) = \sqrt{\sigma} a (nL).  
 \label{eq.19} 
 \end{equation}
This relation is a manifestation of the scaling property of the SSE in
the case $m=0$.

Another definition of the constituent mass, denoted by $\tilde{\mu}_0$,
was used in Refs.~\cite{ref.1,ref.3} in the so called ``einbein
approximation" (EA) where the second extremum condition in
Eq.~(\ref{eq.10}) is written not for the operator $H^{(1)}_{\rm R}$ but for
the eigenvalues $M_0( nL)$. A priori it is not clear whether in both
definitions the extremal values $\mu_0(nL)$ and $\tilde{\mu}_0(nL)$
coincide or not, therefore let us compare them. In the EA
Eq.~(\ref{eq.16}) is rewritten as

\begin{equation}
 \left[ \frac{\vec{p}^{\,2} + m^2}{\tilde{\mu} (t)} + \sigma r \right]
 \tilde{\psi} = \varepsilon (nL) \tilde{\psi} ,
 \label{eq.20}
\end{equation} 
with 
\begin{equation}
 \varepsilon (nL) = \left( \frac{\sigma^2}{\tilde{\mu}} \right)^{1/3}
 A(nL), 
\label{eq.21} 
\end{equation}
i.e., it reduces to the Airy equation with $\varepsilon (nL) = M_0(nL)
- \tilde{\mu}$ and the quantities A(nL) in Eq.~(\ref{eq.21}) are the
zeros of the Airy function. The  constituent mass $\tilde{\mu}_0(nL)$
is now determined by the condition
\begin{equation}
  \frac{d \varepsilon (\tilde{\mu})}{d \tilde{\mu}} + 1 = 0,
 \quad (m = 0).  
\label{eq.22} 
\end{equation}
Then from Eqs.~(\ref{eq.21}, \ref{eq.22}) one obtains that
\begin{equation}
 \tilde{\mu}_0 (nL) = \sqrt{\sigma} \left( \frac{1}{3} A (nL)
 \right)^{3/4} = \sqrt{\sigma}  \tilde{a} (nL). 
 \label{eq.23}
\end{equation}
To compare $\mu_0(nL)$ and $\tilde{\mu}_0(nL)$ one can use the numbers
presented in Appendix \ref{sect.A.1} (see Tables \ref{tab.A.1} and
\ref{tab.A.2}) from which the corresponding universal numbers $a(nL)$
and $\tilde{a}(nL)$ can be determined.

The largest difference between $\mu_0(nL)$ and $\tilde{\mu}_0(nL)$ was
found for $S$ waves and is increasing with growing radial quantum
number $n$ from 5\% for the $1S$ state to 7\% for the $5S$ state.
However, this difference is falling with increasing $L$, being only
1.7\% for $L=5\,(n=0)$. So, $\mu_0$ and $\tilde{\mu}_0$ are
numerically very close. In contrast to the eigenvalues $M_0(nL)$ for
the Salpeter and Airy equations a large difference is found between
some matrix elements (m.e.) like $<1/r^3>$ (for any $L\neq 0$) which
define the fine-structure splittings.  This difference can be as large
as 30-50\% in some cases (see Tables \ref{tab.A.1}, \ref{tab.A.2}).
Moreover, while for the SSE these m.e. are growing, they are slightly
decreasing for the Airy equation. It is worth to notice that these
differences between the m.e.  would be  much larger if a fixed
constituent mass, as in potential models, would be used.

The reason behind such discrepancies may be connected with the
different asymptotic behavor of the wave functions(w.f.). For the SSE
Eq.~(\ref{eq.15}) it falls as $\exp(-\surd\sigma\,r)$ \cite{ref.18}
while for the Airy Eq.~(\ref{eq.20}) the w.f. decreases as
$\exp(-\sqrt{\tilde{\mu}_0 \sigma}\,r^{3/2})$.  Therefore the definition
(\ref{eq.16}) of the constituent quark  mass as well as the
calculations of the m.e. with the use of the unperturbed Hamiltonian
$\tilde{H}^{(1)}_{\rm R}$ has to be considered as preferable compared to the EA.

Note a useful relation between the m.e.:
\begin{equation}
 \langle \sigma r \rangle = 2 \mu_0 (nL)
 \label{eq.24} 
\end{equation}
and 
\begin{equation}
 \langle 1/ r \rangle = \sqrt{\sigma} \langle 1/\rho \rangle_{nL},
 \label{eq.25} 
\end{equation} 
where $\langle 1/\rho \rangle$ is independent of $\sigma$  but does
depend on the quantum numbers.
 
It is worthwile to discuss some common features and differences between
the QCD string approach used here and the RPM which was an essential
step in our understanding of hadronic spectra.

First of all we have shown that the Hamiltonian used in the RPM is more
than a model one, as it can be deduced from the meson Green's function
in QCD for not too large angular momenta $L$, assuming that the string
correction coming from the part $\Delta H_{\rm str}$ Eq.~(\ref{eq.9})
is neglected or considered as a perturbation.

However, in the QCD string approach the mass $m$ must be the current
quark mass, since just the current mass enters the meson Green's
function in the FFS representation, in particular for light quarks
$m=0$ and for the strange quark $m_s = 140-160$ MeV are taken
\cite{ref.3}, while in the RPM $m=(m_s + m_d)/2 = 220$ MeV, $m_s= 465$
MeV are taken in \cite{ref.13}.

Since in the QCD string approach the Hamiltonian as well as the string and
self-energy corrections are calculated with the use of just the same FFS
representation the whole  picture is simplified and the spin-averaged
mass can be expressed through the only parameter--the string tension
$\sigma$.

The constituent mass of a quark does not enter the final form of the
Hamiltonian Eq.~(\ref{eq.15}) and the notion of the constituent mass appears to be
necessary only when one takes into account the spin-dependent
interaction and also the string and self-energy corrections to the
meson mass. In the FFS representation all these terms are inversely
proportional to just the same auxiliary  field $\mu$ \cite{ref.1} which is strictly
detremined from the extremum condition due to the procedure of
canonical quantization and appears to be the kinetic-energy operator.
However,to calculate these corrections the variable $\mu$ must not be
used as an operator and we have defined the constituent mass as the
expection value of the quark kinetic energy. This definition of the
constituent mass is in accord with another one--the variational
definition of the constituent mass used in Ref.~\cite{ref.3}

It is important that the constituent mass in our case depends on the
quantum numbers and increases with growing $L$ and $n_r$.

Finally, instead of the string and self-energy corrections considered
in next Sections (they are negative) in the RPM  a universal negative
constant is introduced.

\section{The string correction and the slope of the Regge trajectory}
\label{sect.6.0}

It is known that for the Salpeter equation (\ref{eq.16}) (or for the
unperturbed Hamiltonian $\tilde{H}^{(1)}_{\rm R}$) the squared masses $M_0^2(nL)$ can
be approximated (with an accuracy of about 1\% for $L\neq 0$) by the
``string formula'' \cite{ref.19},
\begin{equation}
 M^2_0 ({\rm approx}) = 8\sigma L + 4 \pi \sigma (n + 3/4).
 \label{eq.26} 
\end{equation}
The exact values of $M_0^2(nL)$ together with those of $M^2_0$(approx) ($
L\leq 6$ , $n=0$) are given in Table~\ref{tab.1.0} from  which one can
see that the differences between them  are indeed $\leq$  1\% for $L
\geq  2$.

\begin{table} 
\caption{The masses $M_0^2(L)$ and $M_0^2 ({\rm approx}) = 8 L + 3\pi\sigma$
 for the ground states ($n=0$) with $L\leq 6$ ($\sigma = 0.18$
 GeV${}^2$).}
\label{tab.1.0} 
\begin{tabular}{|l|r|r|r|r|r|r|r|} 
\hline
 $L$  &      0 &    1 &      2 &    3 &      4&  5  &       6 \\
\hline
 $M^2_0(nL)$ & 1.7940 & 3.2126 & 4.6431 & 6.0777 & 7.5142 & 8.9518 &
 10.3900\\
 $8L\sigma + 3\pi\sigma$ & 1.696  & 3.1365 & 4.5765 & 6.0165 & 7.4565 &
 8.8965 & 10.3365 \\
 difference  &   5.4\% & 2.3\%   &   1.4\% &  1.0\%  &
 0.77\% &  0.62\%  &   0.5\% \\ 
\hline
\end{tabular} 
\end{table}

As is clear from the approximation (\ref{eq.26}), 
the slope of the Regge trajectory  for the SSE is $(8\sigma)^{-1}$,
i.e. $\alpha'_L = 0.69$ GeV${}^{-2}$ for  $\sigma  =0.18$ GeV${}^2$,
which is 17\% smaller than the experimental number Eq.~(\ref{eq.3}),
$\alpha'_L = 0.81 \pm 0.01$ GeV${}^{-2}$.  Note that the string
corrections  which come from the term Eq.~(\ref{eq.9}) are also
proportional to $L$ and therefore affect the Regge slope. The
situation appears to be different in two domains: $L \leq 4$ and $L\geq
5$ respectively, and we consider them separately.

\subsection{Case A. $L\leq 4$}
\label{sect.6.1}

By the definition (\ref{eq.9}) $\Delta H_{\rm str}$ gives a
negative correction to the eigenvalues $M_0(nL)$; its magnitude turns
out to be relatively small,  $\sim -100$ MeV, and therefore this term
can be considered as a perturbation \cite{ref.11}.
\begin{equation}
  \Delta_{\rm str} ( nL ) = \langle \Delta H_{\rm str} \rangle =
 - \frac{\sigma L(L+1)}{\mu_0 (nL)}\left\langle \frac{1}{r (6 \mu_0 + \sigma r)}
 \right\rangle.
 \label{eq.27}
\end{equation}
In Eq.~(\ref{eq.27}) we have used that the integral $\int^1_0 d\beta
(\beta - 1/2)^2$ is equal to 1/12 and the operators $\nu$ and $\mu$
were replaced by their extremal values Eq.~(\ref{eq.12}) and
Eq.~(\ref{eq.14}). The factor in brackets can also be
approximated  (with an accuracy better than 3\%) replacing $\sigma r$
by $<\sigma r>$. Then the string correction is
\begin{equation}
  \Delta_{\rm str} ( nL ) = - \frac{\sigma L(L+1) \langle 1/r \rangle}
 {\mu_0 (6 \mu_0 + \langle \sigma r  \rangle)} .
 \label{eq.28}
\end{equation}
Due to the relations (\ref{eq.18}) and (\ref{eq.24}) for the linear
potential the correction $\Delta_{\rm str}$ becomes
\begin{equation}
  \Delta_{\rm str} = - \frac{\sigma L(L+1) \langle 1/r \rangle}{8 \mu^2_0}
 = -  \frac{2 \sigma^{3/2} \langle 1/\rho \rangle
 L(L+1)}{M^2_0} . 
 \label{eq.29} 
\end{equation}

Note that in Eq.~(\ref{eq.29}) the m.e. $<1/\rho >\sqrt{L+1}$ is
almost constant, varying from 0.787 for $L=1$ to 0.741  for $L=4$ (see
Table~\ref{tab.A.1}).  The values of $\Delta_{\rm str}$ (using the m.e.
$<1/r>$ from Tables~\ref{tab.A.2} and \ref{tab.A.3} ) are given in
Table~\ref{tab.2.0}.

\begin{table} 
\caption{The string corrections $\Delta_{\rm str}$ in MeV and the mass
 $M_0(L)$ in GeV, for the ground states ($L \leq  6$).} 
\label{tab.2.0}
\begin{tabular}{|l|r|r|r|r|r|r|} 
\hline
 $L$    &        1    &      2  &       3 &        4 &         5 & 6 \\
\hline
 $M_0(L)$   & 1.7924 & 2.1549 &  2.4653   & 2.7412   & 2.9920
 &  3.2234 \\
 $\Delta_{\rm str}(L)$ & -52.9 &  -86.9 & -113.0 &  - 132.7 &
 -153.7 &  - 170.7 \\
 $\Delta_{\rm str}({\rm asym})^a$ & $-\;$ & $-\;$ & $-\;$
 & -142.4 & -182.9 & -219.1 \\ 
\hline
\end{tabular} 

  ${}^a$) For $L \leq 3$ the asymptotic formula Eq.~(\ref{eq.37}) is
  not applicable.  
\end{table}

For comparison in Table~\ref{tab.2.0} the string corrections valid for
large $L$ $(L \geq 5)$ (see the asymptotic string correction formula
Eq.~(\ref{eq.38})) are also given.

\begin{table} 
\caption{The squared masses $M^2=(M_0 + \Delta_{\rm str})^2$ in GeV${}^2$ for 
the ground states ($L\leq 6$, $\sigma  =0.18$ GeV${}^2$).} 
\label{tab.3.0}
\begin{tabular}{|l|r|r|r|r|r|r|} 
\hline
   L &            1   &       2  &       3   &     4   &     5  & 6\\
\hline
 $(M_0+M_{\rm str})^2$ & 3.026 &    4.277   &   5.533  &  6.794
&  8.0567 &
 9.319\\
 $M^2_{\rm str}$(asym)${}^a$ & $-\;$ & $-\;$ & $-\;$ &
 6.754 & 7.891 & 9.026\\ 
\hline
\end{tabular}

${}^a$) see the footnote to Table \ref{tab.2.0}
\end{table}

Now one can analytically calculate the Regge slope for the ``corrected " mass:
 
\begin{equation}
 M(nL) =   M_0(nL)   + \Delta_{\rm str}(nL)\;   (L \leq  4),
 \label{eq.30}
\end{equation}
then the squared mass
\begin{equation}
 M^2(L) = M_0^2(L)  - \frac{4\sigma^{3/2} L(L+1) \langle 1/\rho \rangle}{M_0}
 +  \Delta^2_{\rm str} .
 \label{eq.31}
\end{equation}
If one neglects $\Delta^2_{\rm str}$ in Eq.~(\ref{eq.31}) which is
small ($\leq 0.016$ GeV${}^2$ for $L\leq 4$) and uses the approximation
(\ref{eq.26}) for $M_0^2 (L)$ then for the orbital excitations with
$n = 0$ the squared mass Eq.~(\ref{eq.31}) becomes
\begin{equation}
 M^2 (L) = 8\sigma L - \sigma \frac{\surd 2 \langle 1/\rho \rangle
 (L+1)} {\sqrt{L + 3\pi/8}} + 3\pi\sigma = \left( \alpha'_L
 \right)^{-1} L +  3\pi\sigma
 \label{eq.32} 
\end{equation}
where the inverse Regge slope in Eq.~(\ref{eq.32}) is
\begin{equation}
 \left( \alpha'_L \right)^{-1} = \left(8 -\frac{\surd 2
\langle 1/\rho \rangle (L+1)}
 {\sqrt{L + 3\pi/8}} \right) \sigma = (6.95 \pm 0.02) \sigma.
 \label{eq.33} 
\end{equation}

The values of $(\alpha'_L)^{-1}$ are practically constant, see the
numbers in Table~\ref{tab.A.3}, varying from the value $6.930 \,\sigma$
for $L=1$ to $6.970 \,\sigma$ for $L=4$ and we take here
$(\alpha'_L)^{-1} = (6.95 \pm 0.02) \,\sigma$. Then
\begin{equation}
 M^2(L) = 6.95 \sigma  L  +  3\pi  \sigma 
 \label{eq.34}
\end{equation}
or
\begin{equation}
   L  =   0.144 M^2(L)/\sigma  - 1.358 .  
 \label{eq.35}
\end{equation}
It gives for $\sigma  = 0.18$ GeV${}^2$  the Regge slope
\begin{equation}
 \left( \alpha'_L \right)^{-1} = 1. 25\, {\rm GeV}^2\; {\rm or} \;
  \alpha'_L = 0.80 \, {\rm GeV}^{-2},
 \label{eq.36}
\end{equation}
in good agreement with the  experimental number given in Eq.~(\ref{eq.3})
$\alpha'_L ({\rm exp}) = 0.81 \pm 0.01$ GeV${}^{-2}$. Thus, due to the string corrections we have obtained the
correct Regge slope for the spin-averaged masses. However, the intercept
in Eq.~(\ref{eq.35}) has a very large magnitude and an additional contribution
to the meson mass must be taken into account. We discuss this contribution
in Sect.~\ref{sect.7.0}

\subsection{Case B. Large $L$}
\label{sect.6.2}

For large $L$ the extremal value of the operator $\nu$ is not equal to
$\sigma r$ but turns out to depend on the parameter $\beta$ as well as
on the operator $\mu(t)$. In this case it is a difficult problem to
find the exact eigenvalues $M$(asym) of the Hamiltonian $H_{\rm R}$,
therefore in Ref.\cite{ref.11} the eigenvalues of $H_{\rm R}$ have been
calculated in the quasiclassical approximation with the following
result,
\begin{equation}
 M^2 ({\rm asym}) = 2\pi\sigma\sqrt{L(L+1)} + 3\pi\sigma.
 \label{eq.37} 
\end{equation}
Here, in the asymptotic mass formula (\ref{eq.37}) the string correction is already
taken into account and the constant $3\pi\sigma$ is kept to match the
solutions for large $L$ to those for $L\leq 4$. Now, for comparison one can formally
define the string correction for large $L$ as the difference between
the asymptotic mass Eq.~(\ref{eq.37}) and the unperturbed mass $M_0(nL)$
Eq.~(\ref{eq.18})
\begin{equation}
 \Delta_{\rm str} ({\rm asym}) = \sqrt{3\pi\sigma +
 2\pi\sigma\sqrt{L(L+1)}} - M_0 (L), \; (L >> 1, n=0).  
 \label{eq.38}
\end{equation}

The asymptotic masses are less than $M_0 (L)$ Eq.~(\ref{eq.16}) for
$L\geq 4$.  The magnitude of $\Delta_{\rm str}$ is increasing with
growing $L$ and for $L=6$ $\Delta_{\rm str} ({\rm asym})$ is already 
$\approx 220$ MeV.

From the numerical values of $\Delta_{\rm str}$ (asym) (see
Table~\ref{tab.3.0}) one can see that for $L= 4$ both string
corrections, from the asymptotic formula Eq.~(\ref{eq.38})and from
Eq.~(\ref{eq.30}), practically coincide and in what follows the
string correction will be taken in the form (\ref{eq.30}) for $L\leq
4$ and from Eq.~(\ref{eq.38}) for $L\geq 5$ ( when the masses $M$(asym)
are smaller, see Table~\ref{tab.3.0}).

For $L >> 1$ the Regge slope in Eq.~(\ref{eq.38}) is
$(2\pi\sigma)^{-1}$, i.e. for $\sigma = 0.18$ GeV${}^2$,
$\alpha'_L(L>>1) = 0.88$ GeV${}^{-2}$ is larger than for $L\leq 4$ and
coincides with $\alpha'_J$ for the $\rho$-tracjectory. Such a picture
is partly seen in experiment, where for $L=5$ the difference $M^2 (a_6)
- M^2 (\rho_5)$ is relatively small and  corresponds to the large
value $\alpha'_L \approx 1.1$ GeV$^{-2}$. However, this growth of $\alpha'_L$
is likely to be connected with another reason--an effective decreasing
of the string tension at large distances due to new channels being
opened. This effect is considered in our paper\cite{ref.6}.

The calculated meson masses (see Table~\ref{tab.3.0}) still are large
compared to experiment and to get agreement between them a negative
constant (a fitting parameter) must be added to the squared mass
$M^2(nL)$\cite{ref.13}. Here we shall not introduce a fitting constant,
but instead take into account the quark self-energy correction to the
meson mass.

\section{The quark self-energy contribution and meson masses}
\label{sect.7.0}

Recently it was observed that a negative constant must added to
the meson mass, which comes from  the nonperturbative quark self-energy
contribution created by the color magnetic moment of the quark
\cite{ref.14}. This constant is rather large and  was calculated  with
the use of the Fock-Feynman-Schwinger representation of the quark Green's
function. The total nonperturbative self-energy contribution, both from
the quark and the antiquark, was found to be fully  determined by  the string
tension and by the current mass (flavor) of the quark:
\begin{equation}
 \Delta_{\rm SE} (nL) = - \frac{4\sigma\eta(f)}{\pi\mu_0 (nL)}.
 \label{eq.39}
\end{equation}
Here $\mu_0(nL)$ is just the constituent mass defined by
Eq.~(\ref{eq.17}). The constant $\eta(f)$ depends on the flavor: its
numerical value for a quark of arbitrary flavor was calculated in
Ref.~\cite{ref.14}, in particular for the light mesons we take as in
Ref.~\cite{ref.14}
\begin{equation}
 \eta(n\bar{n}) = 0.90 .
 \label{eq.40}
\end{equation}
The self-energy terms, as well as the meson masses, are given in
Table~\ref{tab.4.0} for the ground states $(n=0, L\leq 5)$ from which
one can see that $\Delta_{\rm SE} (L)$ decreases as a function of $n$
and $L$, being proportional to $\mu^{-1}_0(nL)$.  Still it is rather
large (equal to -300 MeV) even for $L=5$.

With the self-energy and the string corrections taken into account the
spin- averaged meson mass $\bar{M}(nL)$ is fully determined. The
Coulomb correction will be discussed in the next Section and calculated
in Appendix ~\ref{sect.B.1}.

\begin{table} 
\caption{The nonperturbative quark self-energy correction $\Delta_{\rm SE} (L)$
 and the meson masses $\bar{M}(L)$ in GeV for the ground states 
 ($\sigma = 0.18$ GeV${}^2$,  $\eta = 0.9$).}
\label{tab.4.0}
\begin{tabular}{|l|r|r|r|r|r|r|}
\hline
 $L$               &          0   &   1    &   2 &      3 &     4  &  5  \\
\hline
 $\Delta_{\rm SE} (L)$ & -0.616 & -0.460 & -0.383 & -0.335 & -0.301 & -0.294 \\
 $\bar{M}(0,L)$ &  0.723 &  1.279 &  1.685 &  2.017 &  2.30  & $2.514^a$  \\
 $\bar{M}_{\rm exp}(0,L)$ &  0.612 & $1a_J$(1.252) & $\pi_2$(1.67) &
 $a_4$(2.014) & $\rho_5$ & $a_6 (2.45\pm0.13)$ \\
 & & $1f_J$(1.245)& $\rho_3 (1.69)$ & $f_4$(2.034) &
 $2.33\pm0.04$ & $f_6 (2.47\pm0.050$)  \\
&  & & $\omega_3$(1.67) & & &  \\
\hline
\end{tabular}

 ${}^a$) this mass was calculated from the asymptotic formula  (\ref{eq.37}).
\end{table}

The meson mass is now given by 
\begin{equation}
 \bar{M} (nL) = M_0 (nL) - \frac{\sigma \langle 1/r \rangle L(L+1)}
 {\mu_0 ( 6\mu_0 + \langle \sigma r \rangle)}
 - \frac{4\sigma\eta}{\pi\mu_0}, \; (L \leq 4)
 \label{eq.41}
\end{equation}
and for the linear potential can be written as
\begin{equation}
 \bar{M} (nL) = M_0 (nL) - \frac{2 \sigma \langle 1/r \rangle L(L+1)}
 {M^2_0} - \frac{16\sigma\eta}{\pi M_0},
 \label{eq.42}
\end{equation}
using the relations (\ref{eq.18}) and (\ref{eq.24}).
The calculated meson masses $(L\leq 4 )$ coincide with good accuracy 
with the experimental values ( see Table~\ref{tab.4.0} ).

For large $L$ $(n=0)$
\begin{equation}
\bar{M} = \sqrt{3\pi\sigma + 2\pi\sigma\sqrt{L(L+1)}} + \Delta_{\rm SE} (L).
 \label{eq.43}
\end{equation}

\section{The intercept of the Regge trajectory}
\label{sect.8}

From the mass formula (\ref{eq.42}) it follows that the self-energy
term enters $\bar{M}(nL)$ in such a way that the negative constant
$C_0$,
\begin{equation}
 C_0 = - \frac{32 \sigma\eta}{\pi},
 \label{eq.44} 
\end{equation}
appears in the squared  spin-averaged  mass $\bar{M}^2(L)$:
\begin{equation}
 \bar{M}^2 (nL) = (M_0 + \Delta_{\rm str} )^2 
 - \frac{32 \sigma\eta}{\pi} + \left(\frac{16\sigma\eta}{\pi M_0} \right)^2 .
 \label{eq.45}
\end{equation}
Here the terms $\Delta^2_{\rm str}$ and $\Delta_{\rm str} \,
\Delta_{\rm SE}$  will be neglected, because they give small contributions
for $L\leq 4$, while the term $\Delta^2_{\rm SE}$ is kept, since it is
not small in all states. The constant $C_0$ is rather large and for
$\sigma = 0.18$ GeV${}^2$ is equal to $-1.65$ GeV. Using the expression
(\ref{eq.31}) for the mass $(M_0 + \Delta_{\rm str})^2$ and 
Eq.~(\ref{eq.26}) for $M^2_0$, Eq.~(\ref{eq.45}) can be presented as
\begin{equation}
  \bar{M}^2 (L) = \left(\alpha'_L \right)^{-1} L + b(L),
 \label{eq.46} 
\end{equation} 
with 
\begin{eqnarray}
 b(L) & = & \sigma \left[ 3\pi - \frac{32\eta}{\pi} +
 \frac{32\eta^2}{\pi^2(L + 3\pi/8)} \right], \, (L \neq 0), \nonumber \\
 b(L = 0)
 & = & \sigma
 \left[ 3\pi - \frac{32\eta}{\pi} + \frac{256\eta^2}{\pi^2 M^2_0
 (1S)} \right],
 \label{eq.47}
\end{eqnarray}
where for $M_0(L=0)$ it is better to use the exact value, 
$M_0(1S) = 3.157\,\surd \sigma$ and $\alpha'_L$ was already defined 
by the expression (\ref{eq.33}). From (\ref{eq.46}) the intercept is
\begin{equation}
 \alpha_L(0)  = \alpha_L(M^2=0) = - \alpha'_L\, 
  b(L=0) = - \frac{b(L=0)}{6.95\,\sigma} . 
 \label{eq.48} 
\end{equation}
Note that in $b(L)$ the combination $(3\pi- 32\eta/\pi)\sigma$ is a
small number (equal 0.046 GeV${}^2$ for $\sigma = 0.18$ GeV${}^2$) and
therefore for the intercept the contribution of the self-energy term
$\Delta^2_{\rm SE}$ is dominant.

From  Eq.~(\ref{eq.47}) it is clear that $b(L)$ is sensitive to
the value of the flavor factor $\eta$, which may introduce an
uncertainty on the order of 5\%.

With the use of the  analytical expression (\ref{eq.47})
and the exact value of $M_0(L=0)$, $\eta (n\bar{n}) = 0.90$ the quantity
$b(L=0)$ is equal to
\begin{equation}
 b(L=0)  = 2.365 \,\sigma .
 \label{eq.49}
\end{equation}
Then the intercept given by Eq.~(\ref{eq.48}) takes the value
\begin{equation}
 \alpha_L(0)  = -(\alpha'_L)\, b(L=0) = -2.356/6.95  = -0.34.
 \label{eq.50} 
\end{equation}

This number is in good agreement--larger by 10\% only-- with the
experimental value $\alpha_L(0) = -0.30 \pm 0.02$. It is essential that
the intercept does not depend on the string tension but instead is
sensitive to the flavor parameter $\eta$. Just for this reason the
intercept for the mesons with different flavor depends on the flavor.

So, finally, the Regge $L$-trajectory  calculated in the QCD string
approach with $\sigma = 0.18$ GeV${}^2$ is fully determined,
\begin{equation}
 L  =  0.80 \,\bar{M}^2 (L) - 0.34
 \label{eq.51} 
\end{equation} 
and appears to be very close to Eq.~(\ref{eq.3}) obtained from a fit
to the experimental spin-averaged  meson masses, see Fig.~\ref{fig.1}.
From Eq.~(\ref{eq.51}) the averaged mass $\bar{M}(\pi-\rho)$ is
found:
\begin{equation}
 \bar{M}^2(1S) = 0.425\, {\rm GeV}^2\; {\rm or} \;
 \bar{M}(1S)  =0.652\,{\rm GeV},
 \label{eq.52} 
\end{equation}
which corresponds to a $\pi$-meson mass $\bar{M}(\pi)= 301$ MeV. This
number turns out to be smaller than $M(1S)= 0.723$ GeV calculated
directly  from Eq.~(\ref{eq.30}) and this discrepancy illustrates how
sensitive $\bar{M}(1S)$ is to the approximations used.

\section{Coulomb interaction}
\label{sect.9.0}

In the previous sections good agreement of the spin-averaged meson
masses  (for the ground states with $L\neq 0)$ was obtained without
taking into account the Coulomb interaction. It is of interest to check
whether the Coulomb effects are actually suppressed for $L\geq 0$
states and how large is Coulomb correction to $\bar{M}(\pi-\rho)$.

To this end we solve the Salpeter equation with the string
potential taken as a linear plus Coulomb term, i.e., with the Cornell
potential:
\begin{equation}
 V_C(r) = \sigma r - \frac{4}{3} \frac{\alpha_s}{r},
 \label{eq.53}
\end{equation}
where $\alpha_s =$ constant can be used, since the light mesons have
very large sizes, $R\geq 1.0$ fm, and at such distances the strong
coupling is saturated and close to the ``freezing" value\cite{ref.15}.
If for the string tension one takes $\sigma = 0.19$ GeV${}^2$, then
the fitted value of $\alpha_s$ appears to be just the same as for 
heavy quarkonia\cite{ref.20,ref.21},
\begin{equation}
 \alpha_s = 0.39 
 \label{eq.54} 
\end{equation}
However, the masses of the ground states, including the $1S$ state, can
be nicely described with a smaller value for the coupling constant,
$0.20 \leq \alpha_s \leq 0.39$, if correspondingly the value of
$\sigma$ is taken from the range $0.18$ GeV${}^2 < \sigma \leq 0.19$
GeV${}^2$.  

The main characteristics of the $q\bar{q}$ system like the eigenvalues
$M_C(nL)$ of Eq.(\ref{eq.16}) using the Cornell potential, the
constituent masses $\mu_C(nL)$  defined by Eq.~(\ref{eq.17}) together
with the string and the self-energy corrections are presented in
Appendix~\ref{sect.B.1} in Tables~\ref{tab.B.1} and \ref{tab.B.2}.
Here in Table~\ref{tab.5.0} we give only the results of our
calculations for the spin-averaged masses  $\bar{M}_C(nL)$. Note that
in the Coulomb case the relation (\ref{eq.18}) is not valid and
therefore the meson mass $\bar{M}_C(nL)$ as well as $\Delta_{\rm str}$
and $\Delta_{\rm SE}$ should be written through the constituent mass
(denoted as $\mu_C(nL)$) as in Eq.~(\ref{eq.41}) 
(see Table~\ref{tab.B.1} where the eigenvalues are given for $\sigma =
0.19$ GeV${}^2$, $\eta = 0.90$, and $\alpha_s = 0.39$).

With the use of the string and the self-energy corrections from
Table~\ref{tab.B.2} the spin-averaged meson masses $\bar{M}_C(L)$,
Eq.~(\ref{eq.41}), are determined and their values are given in
Table~\ref{tab.5.0} together with the experimental numbers.

\begin{table} 
\caption{ The spin-averaged masses $\bar{M}_C(L)$ in GeV,
theoretical and experimental, for the ground states (n=0) ($\sigma =
0.19$ GeV${}^2$, $\alpha_s = 0.39$).  }
 \label{tab.5.0}
\begin{tabular}{|l|r|r|r|r|r|r|}
\hline
    $L$          &  0    &  1            &     2         &
    3            & 4       & 5    \\
\hline
 $\bar{M}_C(L)$        & 0.632 &
1.220       & 1.650         & 2.00         & 2.29      & 2.51 \\
 $\bar{M}_{\rm exp}(L)$ & 0.612 &
$\bar{M}(f_J)=1.24$ & $\pi_2(1.66)$ & $a_4(2.014)$ &
  $\rho_3(2.30)$ & $a_6(2.45)$ \\
 & &   &    $\rho_3(1.69)$ & $f_4(2.03)$ &  &   $f_6(2.47)$ \\
 & & $\bar{M}(a_J)= 1.25$ & $\omega(1.65)$ & & & \\
 & & &  $\omega_3(1.67)$ & & & \\ 
\hline
\end{tabular} 
\end{table}

If now one compares the meson masses $\bar{M}_C(L)$ with those for the
linear potential from Table~\ref{tab.3.0}, then one can see that in the
Coulomb case for the $1S$ and $1P$ states a better agreement with the
experimental numbers is obtained, however, in the Coulomb case the
string tension appears to be larger, $\sigma = 0.19$ GeV${}^2$. The
calculated mass $\bar{M}_C(1S) = 0.632$ GeV is very close to the value
Eq.~(\ref{eq.52}) from  the Regge trajectory Eq.~(\ref{eq.50}).

Now the Coulomb correction can be formally defined as the difference
between the exact eigenvalues, $\bar{M}_C(L)$ and $\bar{M}(L)$:
\begin{equation}
 E_C ({\rm exact})=  \bar{M}_C (nL) - \bar{M}(nL)
 \label{eq.55} 
\end{equation}
and compared with the Coulomb corrections $E_C$(pert):
\begin{equation}
 E_C({\rm pert}) = - \frac{4}{3} \alpha_s \langle 1/r \rangle
 \label{eq.56} 
\end{equation}
obtained when the Coulomb interaction is considered as a perturbation
(see Table~\ref{tab.6.0}). In Eq.~(\ref{eq.56}) the m.e.  $<1/r>$  is
to be taken for the linear potential with the same $\sigma$ as in the
Cornell potential.

\begin{table} 
\caption{The exact and perturbative Coulomb corrections
$E_C(L)$ (in MeV)} 
\label{tab.6.0} 
\begin{tabular}{|l|r|r|r|r|r|r|}
\hline
  $L$        & 0     &   1     &   2     &   3   &     4       &  5 \\
\hline
 $E_C$(exact) & -219  &   -132  &   - 103 &   - 86.7 &   - 76.2 &   -
 68.8 \\
 $E_C$(pert)  & -194  &   -126  &   -99.4 &   - 84.9 &   - 75.1
 &   - 68.1 \\ difference  & 11.4\% &    4.5\% &     3.5\%&      2.0\%&
 1.4\% &     1.0\% \\ 
\hline
\end{tabular} 
\end{table}

The numbers in Table~\ref{tab.6.0} demonstrate that the exact and
perturbative corrections coincide with an accuracy better than 5\% for
all states with $L > 0$ (for the $1S$ state the difference is 11\% )
and therefore these corrections can be calculated as a perturbation.

For the $nL$ states one should also take into account the difference between
the exact constituent mass $\mu_C(L)$ and $\mu_0(L)$ for the linear potential;
they are related as follows
\begin{eqnarray}
 \mu_C(1\, L) & \approx & \mu_0(1\, L) +  |E_C| /3 \quad  (n=0), \nonumber \\
 \mu_C(n\, L) & \approx & \mu_0(n\, L) +  |E_C| /4, \quad  (n \neq 0).
 \label{eq.57} 
\end{eqnarray}
This correction to the  constituent  mass is mostly important for the
$1S$ state. For larger $n$ the difference between $\mu_C$ and $\mu_0$
can be neglected. As seen from Table~\ref{tab.6.0}, due to the Coulomb
interaction all masses are shifted down by an amount in the range of 70
to 200 MeV and therefore a larger value of $\sigma$ is needed, $\sigma =
0.19$ GeV${}^2$ for $\alpha_s = 0.39$, than for the linear potential.

However, one cannot take an arbitrary or too large value for $\sigma$,
otherwise the Regge slope $\alpha'_L$ would be
small and in contradiction with the experimental value. Therefore, in
the Coulomb case only values $\sigma = 0.19 \pm 0.10$ GeV${}^2$ are allowed. 
Then to obtain agreement with experiment using $\sigma \leq 0.20$ GeV${}^2$ 
a restriction on the value of the strong coupling constant is found:
\begin{equation}
 \alpha_s \leq 0.40 \quad (\sigma \leq 0.20 \, {\rm GeV}^2),
 \label{eq.58} 
\end{equation}
otherwise correct numbers for the Regge slope and the
intercept cannot be obtained simultaneously.

This upper limit (\ref{eq.58}) for $\alpha_s$ appears to be in accord with the
freezing value of the two-loop $\alpha_B (q^2 = 0) = 0.45$  (with the QCD
constant $\Lambda^{(3)}= 330$ MeV, $N_f = 3$) obtained in background field
theory \cite{ref.15}.

\section{Conclusions}
\label{sect.10.0}

In the framework of the QCD string approach the spin-averaged meson
masses with $L\leq 5\; (n=0)$ have been calculated and expressed
through a single parameter--the string tension $\sigma$ and a set of
universal numbers. In this approach the kinetic energy is of the
same type as in the spinless Salpeter equation. The constituent mass and
the nonperturbative quark self-energy are calculable and also depend on
the string tension only.

This is the first time accurate predictions for the meson masses have
been obtained relying on one parameter only, that is
directly connected to the confinment mechanism in QCD.

The analytical expressions for the slope and the intercept of the Regge
$L$-trajectory (when the spin splittings are not taken  into account)
have been deduced, giving rise to a value $\alpha'_L = (6.95
\sigma)^{-1} = 0.80$ GeV${}^{-2}$ ($\sigma = 0.18$ GeV${}^2$) which
coincides with the experimental number. This $L$-trajectory can be
considered as a universal one since in the approximation of closed
channels it does not depend on spin and isospin .

It is shown that the Regge intercept does not depend on $\sigma$ and
$\alpha (M=0) ({\rm theory})= -0.34$ turned out to be only
10\% larger than $\alpha (M=0) ({\rm exp}) = - 0.30 \pm 0.02$. From
this intercept $\bar{M}(1S) = 652$ MeV corresponds to a $\pi$-meson
mass equal to 300 MeV (chiral effects have been neglected here).

For all orbital excitations with $L \neq 0$ the calculated masses are in a
good agreement with  existing experimental data.

In order to obtain this good agreement with the data we find it necessary
to impose a restriction on the value of $\alpha_s$ that is in accord
with the freezing picture.

\vspace*{10mm}

The authors would like to express their gratitude to the theory group of
Thomas Jefferson National Acelerator Facility (TJNAF) for their
hospitality. This work was partly supported by the RFFI - grant
00-02-17836 and INTAS grant 00-00110.

\appendix

\section{Detailed spectra}
\label{sect.A.1}
 
The eigenvalues and the wave functions of the SSE equation were calculated 
with the help of the code used before \cite{ref.21,ref.22}. The eigenvalues
and relevant matrix elements are given in Tables~\ref{tab.A.1} and
\ref{tab.A.2} for the linear potential and in 
Tables~\ref{tab.B.1}-\ref{tab.B.2} for the Cornell potential.

From Table~\ref{tab.A.2} one can see that the difference between the
m.e. $<1/r^3>$ for the SSE and the Airy equations  for the P-wave
states turn out to be large reaching 40\% for $n\geq 2$. As briefly
discussed in Sect.~\ref{sect.3.0} the reason behind these differences
lies in the different asymptotic behaviors of the eigenfunctions of
these two equations.  In Table~\ref{tab.A.3} we also give the
constituent masses Eq.~(\ref{eq.16}) and the m.e.  $<1/r>$ and
$<1/r^3>$ for the ground states $(n=0)$ with $L\leq 6$.

\begin{table} 
\caption{The eigenvalues $M_0(nS)$, constituent masses
 $\mu_0(nS)$, $\tilde{\mu}_0(nS)$, and matrix elements $<1/r>$
 (in GeV ) for the Salpeter Eq.~(\ref{eq.16})and the Airy equation
 Eq.~ (\ref{eq.20}) with the linear potential $\sigma r$ ($\sigma =
 0.18$ GeV${}^2$, $L=0$).  } \label{tab.A.1}
\begin{tabular}{|l|r|r|r|r|r|} 
\hline
    n      &  0   &    1 &       2  &         3 &       4    \\ 
\hline
 $M_0(nS)$ &   1.3394   &   1.9980 &   2.4985  &  2.9151 & 3.2797   \\
 $\mu_0(nS)$ &  0.3348   &   0.4995 &   0.6246 &   0.7289 &  0.8199 \\
 $\tilde{\mu}_0 (nS)$ &  0.3519     & 0.5351   & 0.6703 &   0.7826  &
 0.8807(3)  \\ $<1/r>$ (SSE)   & 0.3638    &  0.3299   & 0.2959  &
 0.2734   & 0.2559(5) \\ $<1/r>^a$ (EA)  & 0.3328  &    0.2669 &
 0.2334&   0.2118 &   0.1996(5) \\ 
\hline 
\end{tabular}

 ${}^a$) These m.e. are calculated from the Eq.(18) with $\tilde{\mu}_0(nS)$
 defined by Eq.(20).	  
\end{table}

\begin{table} 
\caption{The matrix elements $<1/r^3>$ (GeV${}^{-3}$), mass eigenvalues
 $M_0(nP)$ (GeV), and constituent masses $\mu_0(nP)$ and
 $\tilde{\mu}_0(nP)$ (GeV) for the P-wave states ($\sigma = 0.18$ GeV${}^2$).
}
\label{tab.A.2}
\begin{tabular}{|l|r|r|r|r|r|}
\hline
  $n$     &     0    &     1   &         2     &       3    &       4       \\
\hline
 $M_0(nP) $  &  1.7924 &    2.3153   &     2.7505  &    3.1291&        3.4682 \\
 $\mu_0(nP)$ &  0.4481  &   0.5788    &    0.6876   &   0.7823 &       0.8671 \\
 $\tilde{\mu}_0 (nP)$ & 0.4620  &   0.6115    &    0.7320  & 0.8335 & 0.9278 \\
 $<1/r^3 >$ (SSE) & 0.0264  &  0.0422  &  0.0539   &   0.0635 &      0.0718 \\
 $<1/r^3 >$ (EA)$=\tilde{\mu}\sigma/4$ & 0.0208   &  0.0275  & 0.0329 &   0.0376 &     0.0417  \\
\hline
\end{tabular}

 ${}^a$)  see the footnote to Table~\ref{tab.A.1}.	    
\end{table} 
	    
\begin{table} \caption{The constituent masses $\mu_0$ (GeV) and the matrix
elements $< 1/r^k>$ (GeV${}^k$), (k=1,3), of the SSE for the ground
 states(n=0) ($\sigma = 0.18$ GeV${}^2$, $L \leq  6$).  }
\label{tab.A.3} 
\begin{tabular}{|l|r|r|r|r|r|r|} 
\hline
   L           &  1      &   2      &    3     &   4     &   5      &
   6\\
\hline
 $\mu_0 (0\,L)$& 0.4481  &  0.5387  &  0.6163  &  0.6853 &
 0.7480  &  0.8058 \\ 
$<1/r>$ &     0.2362  &   0.1867 &   0.1589 &
 0.1406&   0.1274 &  0.1173\\
 $<1/\rho> \sqrt{L+1}$ & 0.787   &
 0.762  &   0.742  &   0.741 &   0.736 &  0.732\\
 $<1/r^3>$ &   0.0264
 &  0.0098  &   0.0054 &   0.0035&  0.0026  &  0.0019\\ 
\hline
\end{tabular} 
\end{table}

The calculated m.e. $<1/r>$ is used to obtain the string and
Coulomb corrections, while the m.e. $<1/r^3>$ can be used to calculate
the hyperfine and fine-sructure splittings for the mesons with $L \neq 0$.

From Table~\ref{tab.B.1} one can see that in the Coulomb case the
constituent masses $\mu_C (1S)$ and $\mu_C (1P)$ are larger by 29\% and
10\% respectively, than for the linear potential (see Table~\ref{tab.A.2})
and therefore $\Delta_{\rm SE}$ is smaller for them.

\newpage

\section{Results for the Cornell potential} 
\label{sect.B.1}

In this appendix we present some auxiliary values for the Cornell potential.

\begin{table}[h]
\caption{The eigenvalues $M_C(L)$, the constituent masses $\mu_C (L)$
 and the m.e. $<1/r>$  (in GeV) for the SSE with the Cornell potential
 for the ground states ($n=0$).  ($\sigma = 0.19$ GeV${}^2$, $\alpha_s =0.39$)
}
\label{tab.B.1}
\begin{tabular}{|l|r|r|r|r|r|r|}
\hline
 $L$      &     0  &      1   &     2   &     3   &     4    &    5 \\
\hline
 $M_C(L)$ &  1.157 &    1.710 &   2.111 &   2 446 &   2.740  &   3.005 \\
 $\mu_C (L)$ &  0.415 &    0.496 &   0.580 &   0.656 &   0.710  &   0.745  \\
 $<1/r>_L$ &  0.484 &    0.266 &   0.202 &   0.170 &   0.145  &   0.130 \\
\hline
\end{tabular}
\end{table}

\begin{table}[h]
\caption{The string and self-energy corrections (in GeV) for the SSE with 
 the Cornell potential ($\sigma = 0.19$ GeV${}^2$, $\alpha_s = 0.39$).
}
\label{tab.B.2}
\begin{tabular}{|l|r|r|r|r|r|r|}
\hline
     $L$ &    0   &       1  &     2   &      3  &     4    &    5 \\
\hline
 $\Delta_{\rm str}$ &    0   &    -0.051&  -0.086 &  -0.112 &  -0.075  & -0.068 \\
 $\Delta_{\rm SE}$ &  -0.525&    -0.439&  -0.375 &  -0.332 &   -0.307 & -0.282 \\
\hline
\end{tabular}
\end{table}

\end{document}